\documentclass{ws-procs975x65}            

\begin{document}
\title{Wormholes as a cure for black hole singularities}

\author{Gonzalo J. Olmo$^{1,2}$, D. Rubiera-Garcia$^3$, and A. Sanchez-Puente$^1$}
\address{$^1$Departamento de F\'{i}sica Te\'{o}rica and IFIC, Centro Mixto Universidad de
Valencia - CSIC. Universidad de Valencia, Burjassot-46100, Valencia, Spain.\\  E-mail: gonzalo.olmo@uv.es}
\address{$^2$Departamento de F\'isica, Universidade Federal da
Para\'\i ba, 58051-900 Jo\~ao Pessoa, Para\'\i ba, Brazil}
\address{$^3$Instituto de Astrof\'isica e Ci\^encias do Espa\c{c}o, Universidade de Lisboa, Faculdade de Ci\^encias, Campo Grande, PT1749-016 Lisboa, Portugal}

\begin{abstract}
Using exactly solvable models, it is shown that black hole singularities in different electrically charged configurations can be cured. Our solutions describe black hole space-times with a wormhole giving structure to the otherwise point-like singularity. We show that geodesic completeness is satisfied despite the existence of curvature divergences at the wormhole throat. In some cases, physical observers can go through the wormhole and in other cases the throat lies at an infinite affine distance.
\end{abstract}


\bodymatter

\section{Introduction}

One of the most problematic predictions of Einstein's theory of General Relativity (GR) is the possibility of generating space-time singularities in processes of gravitational collapse. The notion of singularity in gravitational theories is a subtle issue \cite{Geroch:1968ut,Hawking:1971wb,Hawking:1973uf,Wald:1984rg} that, nonetheless, is sometimes simplified identifying them with the blow up of certain scalar or tensorial quantities.  However, space-time singularities are more a conceptual problem than a technical one \cite{Curiel2009}. The difficulties that they entail are more related with our ability to describe the world than with the possibility of obtaining an absurdly large numerical prediction (curvature divergences). In fact, a space-time is regarded as singular when there exist time-like or null geodesics which are incomplete \cite{Senovilla:2014gza}. Incomplete geodesics are those that cannot be extended to arbitrarily large values of their affine parameter in the past or in the future of a given event.

 Given that in a curved space-time freely falling observers, which follow geodesic paths, are the analogous of inertial observers in Minkowski space and that they communicate with one another through light signals, the possibility of incomplete geodesics implies that observers and/or signals can be created or destroyed, limiting in this way our ability to describe physical processes. If in a given region physical observers and/or signals are not well defined, then there is no possibility of describing what physical processes are taking place there.  This is the actual problem with space-time singularities, that a complete physical description is impossible where the space-time breaks down \cite{Hawking:1976ra}.

The singularity theorems \cite{Theo1,Theo2,Theo3,Theo4,Theo5} confirm that space-time singularities are unavoidable in GR once trapped surfaces are formed, which can occur under very reasonable physical conditions. Thus, if a way out of this problem exists, it must lie beyond the domain of GR. In this sense, 
in a series of recent works \cite{or12a,Olmo:2013gqa,Olmo:2015axa}, we have addressed the question of space-time singularities from a different geometrical perspective. Extensions of GR have been considered assuming that the background geometry is not necessarily Riemannian in the sense that metric and affine structures could be a priori independent \cite{Olmo:2011uz}. This geometrical approach is justified because the continuum limit of condensed matter systems with a defected microstructure requires the use of metric-affine geometry for a proper description \cite{Lobo:2014nwa}. Space-time singularities could somehow be interpreted as defects in a hypothetical microstructure and metric-affine geometry could be of help to deal with such defects. This approach is also useful to explore the role that new geometric structures could play at very high energies.

Thus, our starting point will be the formulation of well-known gravity theories beyond GR on top of an {\it a priori} non-Riemannian background. The implementation of this idea is rather simple, as the so-called Palatini formalism does it straightaway. In the Palatini approach, metric and connection are regarded as independent geometric entities, and the variation of the action is carried out independently with respect to those fields. Thus, though the Palatini variation is sometimes regarded in the literature as a trick to derive the field equations of GR, it actually has deep conceptual implications because {\it a priori} it breaks the Riemannian constraint $\nabla_\mu g_{\alpha\beta}=0$ and allows the field equations to determine how the metric and the independent connection relate to each other. In the particular case of GR, the field equations for the connection naturally lead to the compatibility condition $\nabla_\mu g_{\alpha\beta}=0$. In other theories, however, this relation does not hold, which has a deep impact on the resulting dynamics.

A natural question that arises when one deals with an independent connection is how it couples to the matter fields. In order to be as conservative as possible, we limit ourselves to follow the prescriptions imposed by the Einstein equivalence principle, which is very well supported experimentally. For this reason, we assume that only the metric couples directly to the matter, being all the other gravitational fields (the connection in this case) part of the gravity Lagrangian only. This is the same that one does, for instance, in scalar-tensor theories, where the scalar field is part of the gravitational action and does not couple directly to the matter fields. The scalar field and the matter help generate the metric but only the metric acts on the matter fields. In our case, the connection plays an active role in the determination of the metric by inducing nonlinear terms of the matter fields in the metric field equations. It is these nonlinearities that generate new dynamics and modify the solutions at very high energy densities. When the matter fields are absent (vacuum), the field equations boil down to those of GR and the connection coincides with the Levi-Civita connection of the metric\cite{Olmo:2011uz}. Thus, only in scenarios with matter fields do these Palatini theories generate new dynamics.

In what follows, we will summarize the main results found in recent works\cite{Olmo:2015bya,Bazeia:2015uia,Olmo:2015axa} for two different gravity models formulated \`{a} la Palatini and coupled to spherically symmetric matter sources. These models yield  black hole solutions with up to two event horizons, like the Reissner-Nordstrom solution of GR, but with a wormhole at their center. This last property makes them geodesically complete, including the {\it naked} configurations. Full details on the derivation of the field equations and their application in spherically symmetric scenarios can be found in those papers and references therein.

\section{Born-Infeld gravity model}

Inspired by the Born-Infeld approach to electrodynamics, a Born-Infeld-like gravity theory can be written as \cite{Olmo:2013gqa} $S=\frac{1}{\kappa^2\epsilon}\int d^4x \left[\sqrt{-|g_{\mu\nu}+\epsilon R_{\mu\nu}(\Gamma)|}-\lambda \sqrt{-|g_{\mu\nu}|}\right]+S_m[g_{\mu\nu},\psi]$, 
where  vertical bars inside the square-root denote determinant, and $\epsilon$ is a small parameter with dimensions of length squared, which for convenience we write as $\epsilon=-2l_\epsilon^2$.  The connection equation can be solved formally in terms of the Levi-Civita connection of an auxiliary metric whose form is determined by the space-time metric $g_{\mu\nu}$ and a deformation tensor that depends on the matter fields. Thus, the only dynamical equations to solve are those for the metric. Considering as matter source a spherically symmetric Maxwell electric field, one finds that the line element can be written as
\begin{equation}
ds^2=-\frac{A(x)}{\Omega_+}dt^2+\frac{1}{A(x)\Omega_+}dx^2+r^2(x)(d\theta^2+\sin^2\theta d\varphi^2) \ ,
\end{equation}
where $A(x)=1-\frac{2M(x)}{x}$, $M(x)=M_0(1+\delta_1 G[r(x)])$, $dG/dz=\frac{1}{z^4}\frac{(1+z^4)}{\sqrt{z^4-1}}$, $z=r(x)/r_c$, $M_0$ and $r_c$ are constants, $\Omega_\pm=1\pm1/z^4$, and $r^2(x)=\frac{x^2+\sqrt{x^4+4 r_c^4}}{2}.$
From this formula, it is apparent that the area of the $2-$spheres has a minimum of magnitude $A=4\pi r_c^2$ at $x=0$. This signals the presence of a wormhole, which can be further supported by studying the properties of the electric flux. In fact, given that the matter action only considers a free electric field, the wormhole structure confirms that this solution is a geon in Wheeler's sense. Thus, this wormhole is supported by an electric field trapped in the topology, which implies that  $x\in]-\infty,+\infty[$.

\begin{figure}[h]
\begin{center}
\includegraphics[width=0.55\textwidth]{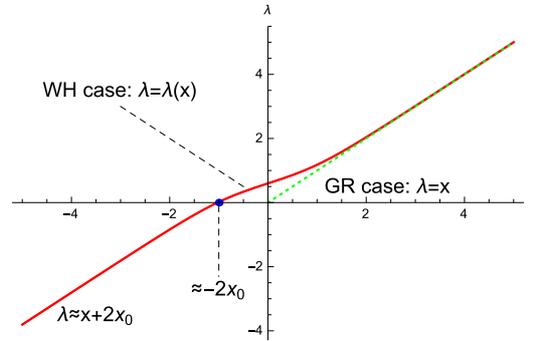}
\end{center}
\caption{Affine parameter $\lambda(x)$ as a function of the radial coordinate $x$ for radial null geodesics (outgoing in $x>0$). In the GR case (green dashed curve in the upper right quadrant), $\lambda=x$ is only defined for $x\ge 0$ (incomplete). For radial null geodesics in our wormhole spacetime (solid red curve), $\lambda(x)$ interpolates between the GR prediction and a shifted straight line $\lambda(x)\approx  x+2x_0$, with $x_0\approx 0.59907$. In this plot $E=1$ and the $x-$axis is measured in units of $r_c$. } \label{Fig:affine_nullradial}
\end{figure}

It is easy to verify that the line element for $r\gg r_c$ recovers the Reissner-Nordstr\"om solution of GR and that the only differences arise for values of order $z=r/r_c\lesssim 3$ (with $r_c=\sqrt{\kappa q l_\epsilon}$, and $q$ representing the electric charge). We just need to focus on the behavior of geodesics in this region to determine if they can be extended across the wormhole. The geodesic equation for the above line element takes the form $\frac{1}{\Omega_+^2}\left(\frac{dx}{d\lambda}\right)^2=E^2-\frac{A(x)}{\Omega_+}\left(\frac{L^2}{r^2(x)}-k\right) \ ,$
where $E$ and $L$ are constants related to the energy and angular momentum of the test particle, and $k=0,-1$ for null and time-like geodesics, respectively. One can verify that time-like geodesics never reach the wormhole, in much the same way as they do in GR, where the singularity is never reached by time-like observers. Null radial geodesics, however, do reach to the singularity in GR and terminate there (null incompleteness). In our case, setting $k=0$ and $L=0$, one finds that the geodesic equation can be exactly integrated to yield $\pm E \cdot \lambda(x)={_{2}{F}}_1[-\frac{1}{4},\frac{1}{2},\frac{3}{4};\frac{r_c^4}{r^4}]  r $ if $ x\ge 0 $, and $\pm E \cdot \lambda(x)=2x_0- {_{2}{F}}_1[-\frac{1}{4},\frac{1}{2},\frac{3}{4};\frac{r_c^4}{r^4}]  r$ if $x\le 0$, 
where $_{2}F_1[a,b,c;y]$ is a hypergeometric function, $x_0={_{2}{F}}_1[-\frac{1}{4},\frac{1}{2},\frac{3}{4};1] =\frac{\sqrt{\pi}\Gamma[3/4]}{\Gamma[1/4]}\approx 0.59907$, and the $\pm$ sign corresponds to outgoing/ingoing null rays in the $x>0$ region. In Fig.\ref{Fig:affine_nullradial}, one can see that these geodesics can be smoothly extended across the wormhole ($x=0$), confirming their completeness. Remarkably, at the wormhole throat there exists a generic curvature divergence which, however, is not an obstacle to having a smooth definition of geodesics everywhere.

\section{Quadratic $f(R)$ gravity}

\begin{figure}
\includegraphics[width=0.45\textwidth]{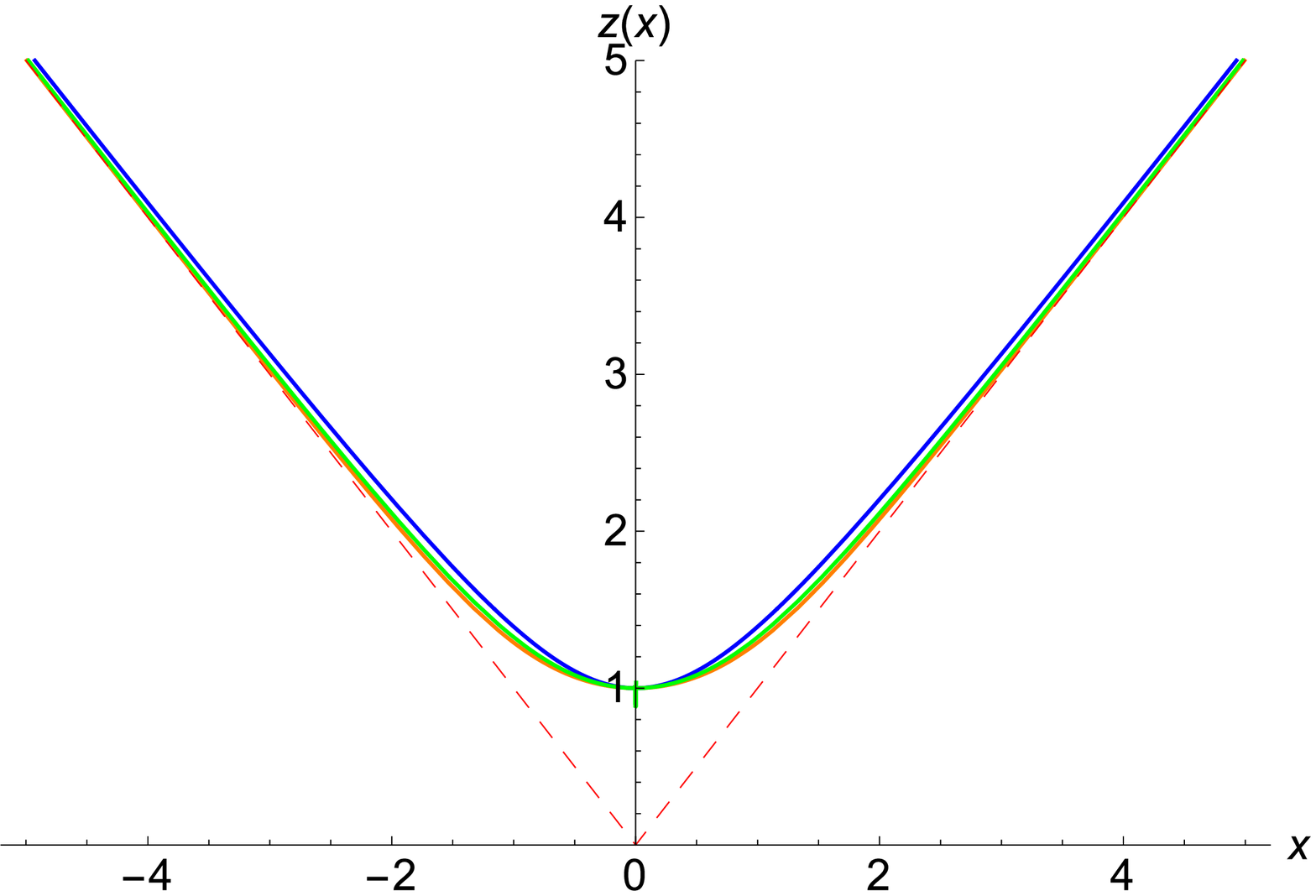} 
\includegraphics[width=0.45\textwidth]{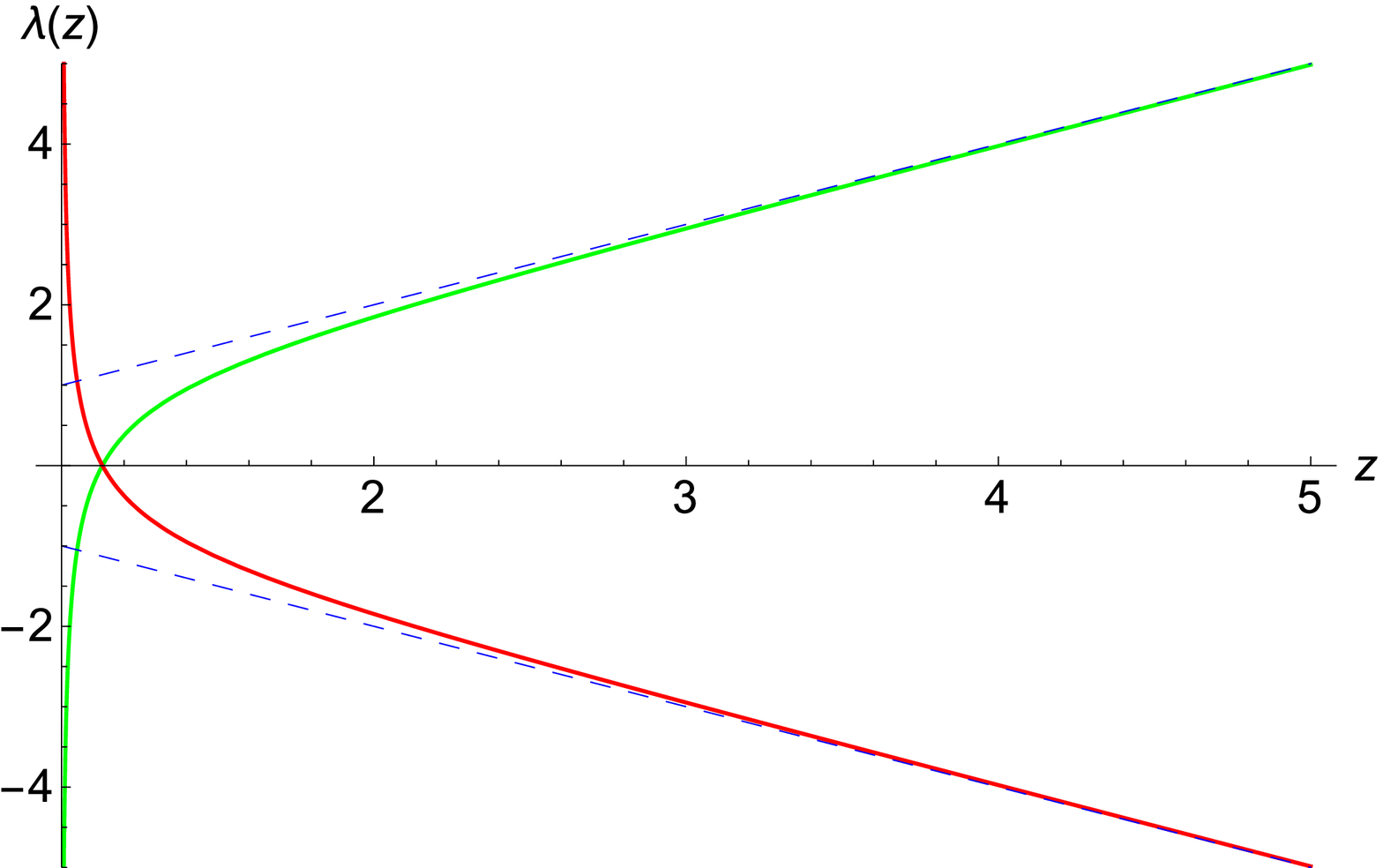}
\caption{(Left) Representation of $r(x)$ in units of  $r_c$ for $\alpha=1/10, 1/2, 4/5$ (blue, green, and orange, respectively). The dashed lines represent $|x|$, the GR case, which is quickly recovered for $x >2 $. (Right) Outgoing (green) and ingoing (red) null radial geodesics with $E=1$ and $\alpha=4/5$. Similar results are found for any other value $0<\alpha<1$.}\label{Fig2}
\end{figure}

Another interesting model, which is perhaps easier to motivate from an effective field theory perspective, is that defined by the gravity Lagrangian $f(R)=R-\lambda R^2$, with $\lambda$ representing some (positive) squared length. In the Palatini approach, $f(R)$ theories also yield modified dynamics by means of nonlinearities induced by the matter terms. In this case, however, the dependence on the matter fields is through the trace of their stress-energy tensor, which prevents the simple case of a Maxwell field. Given that electromagnetic fields have a trace anomaly at the quantum level, we can explore its phenomenological impact by considering nonlinear theories of electrodynamics as matter source. A related example is that of
an anisotropic fluid of the form ${T_\mu}^\nu= \text{diag}[-\rho,-\rho,\alpha \rho ,\alpha\rho]$ with $\alpha$ a constant. The conservation equation implies that $\rho(x)=C/r(x)^{2+2\alpha}$ and, therefore, if $\alpha=1$, this fluid is equivalent to the case of a Maxwell electric field. In general, for this fluid to satisfy the energy conditions, one requires $0\leq \alpha \leq 1$.  Like in the Born-Infeld case, the connection equation can be formally solved in terms of an auxiliary metric, and the physical line element can be taken as \cite{Olmo:2015axa}
\begin{equation}
d{s}^2=\frac{1}{f_R}\left(-A(x)dt^2+\frac{1}{A(x)}dx^2\right)+r^2(x) d\Omega^2 \  \label{eq:ds2} \ ,
\end{equation}
where $f_R=1-1/z^{2+2\alpha}$, $r=r_c z$ and $r_c^{2+2\alpha}\equiv (4\lambda)\kappa^2(1-\alpha)C$. The function $r^2(x)$ is given by solving the relation $x^2=r_c^2z^2(1-1/z^{2+2\alpha})$,
which is plotted in Fig. \ref{Fig2}, showing a wormhole structure.\\\vspace{0.0cm}
Like in the Born-Infeld case shown before, the function $A(x)$ can be solved analytically and its behavior far from the wormhole throat is essentially coincident with the corresponding solution of GR for the value of $\alpha$ chosen. Near the throat, however, the geometry is different, which has an impact on the geodesic structure. The geodesic equation becomes $\left(\frac{dx}{d\lambda}\right)^2=E^2f_R^2-A(x)f_R \left(k+\frac{L^2}{r^2(x)}\right)$.
Here we also find a Reissner-Nordstrom like behavior for time-like trajectories and null ones with $L\neq 0$. For null radial geodesics, however, the behavior is different. The exact analytical solution is of the form $\pm E\tilde\lambda (z)=-\frac{z}{\sqrt{1-z^{-2 (\alpha +1)}}}+2 z \, _2F_1\left(\frac{1}{2},-\frac{1}{2 (\alpha +1)};1-\frac{1}{2 (\alpha +1)};z^{-2 (\alpha +1)}\right)$.
Far away from the throat, we have $\pm E\tilde \lambda\approx z $, like in GR, but near the throat, this expression yields $\pm E\tilde \lambda\approx -\frac{1}{\sqrt{2 \alpha +2} \sqrt{z-1} }=-\frac{1}{|\tilde x|}$ (see Fig. \ref{Fig2}). The divergence of $\lambda(x)$ on both limits, when $z\to \infty$ and when $z\to 1$, implies that null radial geodesics are complete. Interestingly, these wormholes also have curvature divergences at the throat $x=0$, but these results show that they lie beyond the reach of any observer or signal and, therefore, do not belong to the physical space-time.

\section{Summary and conclusions}

In this work we have considered the problem of black hole singularities from the perspective of extended theories of gravity in metric-affine geometries. We have studied two different gravity models and have shown that in both cases the resulting space-times are geodesically complete despite the existence of curvature divergences. This is so for configurations with two event horizons, with one (degenerate) horizon, and with no horizons (naked). The case of naked configurations is particularly relevant because it puts forward that the cosmic censorship conjecture is unnecessary if the space-time topology is nontrivial.

\section*{Acknowledgments}
G.J.O. is supported by a Ramon y Cajal contract, the Spanish grants FIS2014-57387-C3-1-P and FIS2011-29813-C02-02 from MINECO, the grants i-LINK0780 and i-COOPB20105 of the Spanish Research Council (CSIC), the Consolider Program CPANPHY-1205388, and the CNPq project No. 301137/2014-5 (Brazilian agency). D.R.-G. is supported by the FCT grants No.SFRH/BPD/102958/2014 and No.UID/FIS/04434/2013.

\end{document}